\PassOptionsToPackage{unicode}{hyperref}
\PassOptionsToPackage{hyphens}{url}
\PassOptionsToPackage{dvipsnames,svgnames,x11names}{xcolor}
\documentclass[
]{article}
\usepackage{amsmath,amssymb}
\usepackage{lmodern}
\usepackage{iftex}
\ifPDFTeX
  \usepackage[T1]{fontenc}
  \usepackage[utf8]{inputenc}
  \usepackage{textcomp} % provide euro and other symbols
\else % if luatex or xetex
  \usepackage{unicode-math}
  \defaultfontfeatures{Scale=MatchLowercase}
  \defaultfontfeatures[\rmfamily]{Ligatures=TeX,Scale=1}
\fi
\IfFileExists{upquote.sty}{\usepackage{upquote}}{}
\IfFileExists{microtype.sty}{% use microtype if available
  \usepackage[]{microtype}
  \UseMicrotypeSet[protrusion]{basicmath} % disable protrusion for tt fonts
}{}
\makeatletter
\@ifundefined{KOMAClassName}{% if non-KOMA class
  \IfFileExists{parskip.sty}{%
    \usepackage{parskip}
  }{% else
    \setlength{\parindent}{0pt}
    \setlength{\parskip}{6pt plus 2pt minus 1pt}}
}{% if KOMA class
  \KOMAoptions{parskip=half}}
\makeatother
\usepackage{xcolor}
\usepackage{graphicx}
\makeatletter
\def\maxwidth{\ifdim\Gin@nat@width>\linewidth\linewidth\else\Gin@nat@width\fi}
\def\maxheight{\ifdim\Gin@nat@height>\textheight\textheight\else\Gin@nat@height\fi}
\makeatother
\setkeys{Gin}{width=\maxwidth,height=\maxheight,keepaspectratio}
\makeatletter
\def\fps@figure{htbp}
\makeatother
\NewDocumentCommand\citeproctext{}{}
\NewDocumentCommand\citeproc{mm}{%
  \begingroup\def\citeproctext{#2}\cite{#1}\endgroup}
\makeatletter
 \let\@cite@ofmt\@firstofone
 \def\@biblabel#1{}
 \def\@cite#1#2{{#1\if@tempswa , #2\fi}}
\makeatother
\newlength{\cslhangindent}
\newlength{\csllabelwidth}
\newenvironment{CSLReferences}[2] % #1 hanging-indent, #2 entry-spacing
 {\begin{list}{}{%
  \setlength{\itemindent}{0pt}
  \setlength{\leftmargin}{0pt}
  \setlength{\parsep}{0pt}
  \ifodd #1
   \setlength{\leftmargin}{\cslhangindent}
   \setlength{\itemindent}{-1\cslhangindent}
  \fi
  \setlength{\itemsep}{#2\baselineskip}}}
 {\end{list}}
\usepackage{calc}

\ifLuaTeX
\usepackage[bidi=basic]{babel}
\else
\usepackage[bidi=default]{babel}
\fi
\babelprovide[main,import]{american}
\def\languageshorthands#1{}
\ifLuaTeX
  \usepackage{selnolig}  % disable illegal ligatures
\fi
\IfFileExists{bookmark.sty}{\usepackage{bookmark}}{\usepackage{hyperref}}
\IfFileExists{xurl.sty}{\usepackage{xurl}}{} % add URL line breaks if available
\hypersetup{
  pdftitle={BayesEoR: Bayesian 21-cm Power Spectrum Estimation from
Interferometric Visibilities},
  pdfauthor={Peter H. Sims, Jacob Burba, Jonathan C. Pober},
  pdflang={en-US},
  colorlinks=true,
  linkcolor={Maroon},
  filecolor={Maroon},
  citecolor={Blue},
  urlcolor={Blue},
  pdfcreator={LaTeX via pandoc}}

\title{BayesEoR: Bayesian 21-cm Power Spectrum Estimation from
Interferometric Visibilities}

\definecolor{c53baa1}{RGB}{83,186,161}
\definecolor{c202826}{RGB}{32,40,38}

\usepackage[affil-it]{authblk}
\usepackage{orcidlink}
\author[1%
  *%
  ]{Peter H. Sims%
    \,\orcidlink{0000-0002-2871-0413}\,%
    }
\author[2%
  *%
  \ensuremath\mathparagraph]{Jacob Burba%
    \,\orcidlink{0000-0002-8465-9341}\,%
    }
\author[3%
  *%
  ]{Jonathan C. Pober%
    \,\orcidlink{0000-0002-3492-0433}\,%
    }

\affil[1]{School of Earth and Space Exploration, Arizona State
University, USA%
  }
\affil[2]{Department of Physics and Astronomy, University of Manchester,
UK%
  }
\affil[3]{Department of Physics, Brown University, USA%
  }
\affil[$\mathparagraph$]{Corresponding author: %
}
\affil[*]{These authors contributed equally.}
\date{17 January 2024}

\begin{document}
\maketitle

\section{Summary}\label{summary}

\texttt{BayesEoR} is a GPU-accelerated, MPI-compatible Python package
for estimating the power spectrum of redshifted 21-cm emission from
interferometric observations of the Epoch of Reionization (EoR).
Utilizing a Bayesian framework, \texttt{BayesEoR} jointly fits for the
21-cm EoR power spectrum and a ``foreground'' model, referring to
bright, contaminating emission between us and the cosmological signal,
and forward models the instrument with which these signals are observed.
To perform the sampling, we use \texttt{MultiNest}
(\citeproc{ref-buchner:2014}{Buchner et al., 2014}), which calculates
the Bayesian evidence as part of the analysis. Thus, \texttt{BayesEoR}
can also be used as a tool for model selection (see e.g.
\citeproc{ref-sims:2019a}{Sims et al., 2019}).

\section{Statement of need}\label{statement-of-need}

Neutral hydrogen can undergo a spin-flip transition in which the quantum
spins of the proton and electron transition from an aligned to an
anti-aligned state, or vice versa, resulting in emission or absorption
of a photon with a wavelength of 21-cm. The hydrogen 21-cm spin
temperature quantifies the relative number densities of atoms in the
aligned and anti-aligned states. Interferometric 21-cm cosmology
experiments aim to measure the contrast between the 21-cm spin
temperature of neutral hydrogen and the radio background temperature in
the early Universe. By observing this signal at high redshift, we can
learn a wealth of information about the state of the intergalactic
medium during the first billion years of cosmic history. This
information can, in turn, be used to infer properties of the first stars
and galaxies that transformed the hydrogen intergalactic medium from a
cold neutral gas to a hot ionised plasma during the Epoch of
Reionization (EoR). Modern interferometers like
\href{http://reionization.org/}{HERA},
\href{https://www.mpifr-bonn.mpg.de/en/lofar}{LOFAR}, and the
\href{http://www.mwatelescope.org/}{MWA} have been designed to observe
with many antennas simultaneously to maximize their sensitivity to the
21-cm signal from the EoR. These experiments have shown that detecting
this signal is rife with difficulty
(\citeproc{ref-hera:2022}{Abdurashidova et al., 2022};
\citeproc{ref-lofar:2020}{Mertens et al., 2020};
\citeproc{ref-mwa:2020}{Trott et al., 2020}). This is primarily due to
the coupling of bright contaminating sources between us and the
cosmological signal, referred to as ``foregrounds'', with the spectral
structure imparted by the instrument. Existing approaches to recovering
the 21-cm signal from the data lack direct modelling of the observed
covariance between the 21-cm and foreground signals in the data.
Intrinsically, the 21-cm and foreground signals are uncorrelated. The
instrument modulates both signals identically during observation,
however, making them covariant. This covariance can be accounted for by
forward modelling both signals, a key advantage of our approach in
\texttt{BayesEoR}. For a detailed comparison of \texttt{BayesEoR} with
other existing methods, please see section 7.1 of Sims et al.
(\citeproc{ref-sims:2019a}{2019}) and section 1 of Burba et al.
(\citeproc{ref-burba:2023}{2023}).

\begin{figure}
\centering
\includegraphics{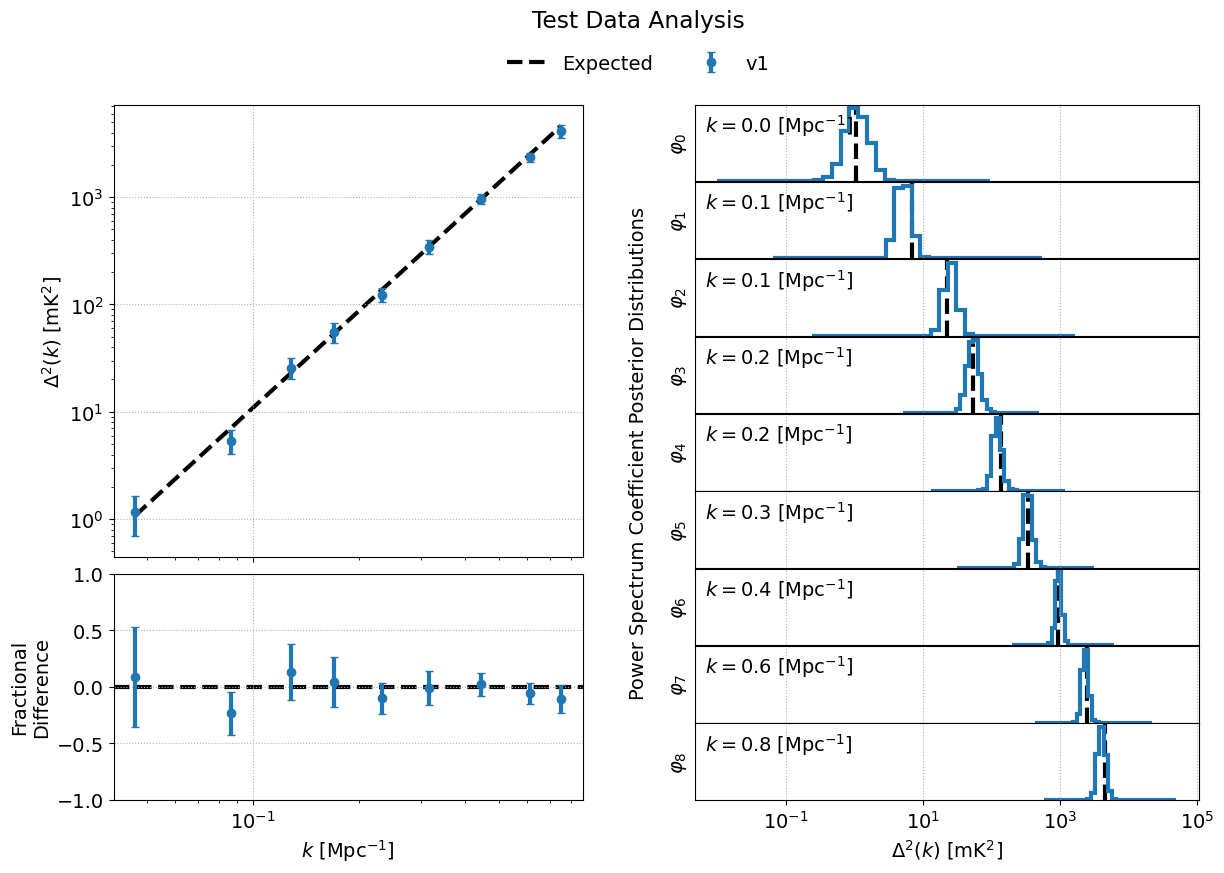}
\caption{Example outputs from a \texttt{BayesEoR} analysis of the
provided test dataset with a known power spectrum. The top left subplot
shows the inferred dimensionless power spectrum estimates with
\(1\sigma\) uncertainties in blue (\(\Delta^2(k)\) as a function of
spherically-averaged Fourier mode, \(k\)) and the expected dimensionless
power spectrum as the black, dashed line. The bottom left subplot shows
the fractional difference between the recovered and expected power
spectra. The right subplots show the posterior distribution of each
power spectrum coefficient in the top left plot (\(\varphi_i\) where
\(i\) indexes the spherically-averaged \(k\) bins) in blue with the
expected power spectrum amplitude as the black, vertical, dashed lines.
\label{fig:example}}
\end{figure}

\texttt{BayesEoR} is a GPU-accelerated, MPI-compatible Python
implementation of a Bayesian framework designed to jointly model the
21-cm and foreground signals and forward model the instrument with which
these signals are observed. Using these combined techniques, we can
overcome the aforementioned difficulties associated with extracting a
faint, background signal in the presence of bright foregrounds.
\texttt{BayesEoR} enables one to sample directly from the marginal
posterior distribution of the power spectrum of the underlying 21-cm
signal in interferometric data, enabling recovery of statistically
robust and unbiased\footnote{Recovery of unbiased estimates of the 21-cm
  power spectrum requires that the field of view of the foreground model
  encompasses the region of sky from which instrument-weighted
  foregrounds contribute significantly to the observed data
  (\citeproc{ref-burba:2023}{Burba et al., 2023}).} estimates of the
21-cm power spectrum and its uncertainties
(\citeproc{ref-burba:2023}{Burba et al., 2023};
\citeproc{ref-sims:2016}{Sims et al., 2016},
\citeproc{ref-sims:2019a}{2019}; \citeproc{ref-sims:2019b}{Sims \&
Pober, 2019}). The power spectrum estimates (\(\Delta^2(k)\)) from an
analysis using the
\href{https://bayeseor.readthedocs.io/en/latest/usage.html\#test-dataset}{test
dataset} and
\href{https://bayeseor.readthedocs.io/en/latest/usage.html\#analyzing-bayeseor-outputs}{plotting
code} provided with \texttt{BayesEoR} can be found in
\autoref{fig:example}. This figure demonstrates the primary output of
\texttt{BayesEoR}: a posterior distribution of the dimensionless power
spectrum amplitude of the 21-cm EoR signal for each spherically-averaged
\(k\) bin in the model (right subplots in \autoref{fig:example}). From
these posteriors, we can derive power spectrum estimates and
uncertainties (top left subplot in \autoref{fig:example}).
Mathematically, the spherically-averaged power spectrum \(P(k)\) is
calculated as \begin{equation}
P(k_i) = \frac{1}{N_{k,i}}\sum_{\mathbf{k}}P(\mathbf{k})
\end{equation} where \(i\) indexes the spherically-averaged \(k\) bins,
the sum is performed over all \(\mathbf{k}\) in a spherical shell
satisfying \(k_i \leq |\mathbf{k}| < k_i + \Delta k_i\), and
\(N_{k, i}\) is the number of voxels in the \(i\)-th spherical shell.
The spherically-averaged dimensionless power spectrum, \(\Delta^2(k)\),
which we infer in \texttt{BayesEoR}, is related to \(P(k)\) via
\begin{equation}
\Delta^2(k_i) = \frac{k_i^3}{2\pi^2}P(k_i)
\end{equation}

\section{\texorpdfstring{Running
\texttt{BayesEoR}}{Running BayesEoR}}\label{running-bayeseor}

Running a \texttt{BayesEoR} analysis requires an input dataset, a model
of the instrument, and a set of analysis parameters. A script is
provided with \texttt{BayesEoR} for convenience which pre-processes a
\texttt{pyuvdata}-compatible dataset
(\citeproc{ref-hazelton:2017}{Hazelton et al., 2017}) of visibilities
per baseline, time, and frequency into a one-dimensional data vector,
the required form of the input dataset to \texttt{BayesEoR}. As part of
the inference, \texttt{BayesEoR} forward models the instrument which
requires an instrument model containing the primary beam response of a
``baseline'' (pair of antennas) and the ``uv sampling'' (the length and
orientation of each baseline in the data). The primary beam response is
passed via a configuration file or command line argument. The uv
sampling is generated by the aforementioned convenience script to ensure
that the ordering of the baseline in the instrument model matches that
in the visibility data vector. Analysis parameters must also be set by
the user to specify file paths to input and output data products and
model parameters used to construct the data model (e.g.~the number of
frequencies and times in the data, the number of voxels in the model
Fourier domain cube, the field of view of the sky model). Note however
that these analysis parameters must be chosen carefully based on the
data to be analyzed (please see section 2.3 of
\citeproc{ref-burba:2023}{Burba et al., 2023} for more details on
choosing model parameters). Accordingly, because \texttt{BayesEoR}
forward models the instrument, we generate a model of the sky as part of
our model visibilities calculation. When the EoR and foregrounds can be
adequately described by a sky model with a field of view equal to the
width of the primary beam, the memory requirements for a
\texttt{BayesEoR} analysis are on the order of 10 GB. This is the case
for the provided test dataset for which the disk and RAM requirements
are \textasciitilde17 GB and \textasciitilde12 GB, respectively. The
field of view for the EoR and foreground sky models can be set
independently, however, which allows for the EoR to be modelled within
the primary field of view of the telescope while the foregrounds can be
modelled across the whole sky. We wish to note however that, in this
fashion, modelling the whole sky can be computationally demanding
depending upon the data being analyzed. For example, in Burba et al.
(\citeproc{ref-burba:2023}{2023}) we show that analyzing a relatively
modest dataset (compared to those typically analyzed by HERA, LoFAR, or
the MWA) can require \textasciitilde400 GB of RAM\footnote{\texttt{BayesEoR}
  uses a routine from the Matrix Algebra for GPU and Multicore
  Architectures (\href{https://icl.utk.edu/magma/}{MAGMA}) library which
  allows for the use of matrices with a memory footprint larger than the
  available RAM on a GPU.}. Please see section 6.1 of Burba et al.
(\citeproc{ref-burba:2023}{2023}) for more details.

\section{Acknowledgements}\label{acknowledgements}

The authors acknowledge support from NSF Awards 1636646 and 1907777, as
well as Brown University's Richard B. Salomon Faculty Research Award
Fund. JB also acknowledges support from a NASA RI Space Grant Graduate
Fellowship. PHS was supported in part by a McGill Space Institute
fellowship and funding from the Canada 150 Research Chairs Program. This
result is part of a project that has received funding from the European
Research Council (ERC) under the European Union's Horizon 2020 research
and innovation programme (Grant agreement No.~948764; JTB). This
research was conducted using computational resources and services at the
Center for Computation and Visualization, Brown University.

\section*{References}\label{references}
\addcontentsline{toc}{section}{References}

\phantomsection\label{refs}

\end{document}